\documentclass[11pt]{article}
\usepackage{amssymb,esvect,amsmath,graphicx,latexsym,amsthm,slashed,eso-pic,hyperref,authblk}

   \DeclareMathOperator{\mev}{MeV}             
    \newcommand{\cI}{{\cal I}}   \newcommand{\cM}{{\cal M}}  \newcommand{\cO}{{\cal O}}   
\newcommand{\ep}{\epsilon}

\newcommand{\pL}{\left(} \newcommand{\pR}{\right)} \newcommand{\bL}{\left[} \newcommand{\bR}{\right]} \newcommand{\cbL}{\left\{} \newcommand{\cbR}{\right\}}  
\newcommand{\beq}{\begin{equation}} \newcommand{\eeq}{\end{equation}}
\newcommand{\bea}{\begin{eqnarray}} \newcommand{\eea}{\end{eqnarray}}

\newcommand{\Eq}[1]{Eq.~(\ref{#1})}

\title{Nuclear Kinetic Equilibrium \\ During Big Bang Nucleosynthesis}
\author[1]{Samuel D.~McDermott}
\author[2]{Michael S.~Turner}
\affil[1]{Fermi National Accelerator Laboratory, Batavia, IL  60510, USA}
\affil[2]{Kavli Institute for Cosmological Physics, Departments of Physics and of Astronomy \& Astrophysics, and the Enrico Fermi Institute, The University of Chicago, Chicago, IL  60637, USA}
\date{\today}                             

\begin{document}
${}$ \hfill {\tt FERMILAB-PUB-18-625-A} \\[-2\baselineskip]
{\let\newpage\relax\maketitle}
\maketitle
\abstract{Sasankan et al.~\cite{Sasankan2018} have recently claimed that there are significant deviations in the distributions of the kinetic energies of nuclei from the Maxwell-Boltzmann form usually assumed in BBN, and further, that these deviations lead to big changes in the predicted light-element abundances.  Solving the relativistic Boltzmann equation perturbatively, we explicitly show that these deviations are not 20\% as claimed, but rather are about $10^{-17}$ in size and hence cannot significantly alter the predicted light-element abundances.  We discuss and compute two related effects:  ${\cal O}$(0.1\%) corrections to the kinetic distributions of nuclei that arise from small relativistic corrections to the MB distribution and a much smaller effect, nuclear kinetic drag, which arises from the heat transferred from the EM plasma to nuclei that is needed to maintain kinetic equilibrium.}

\section{Introduction}
The detailed and correct computation of big-bang nucleosynthesis (BBN) dates back 51 years to the seminal papers of Wagoner, Fowler and Hoyle \cite{WFH}.  Since that time, the BBN code has been updated for numerous physics effects including finite-temperature Coulomb and radiative corrections \cite{RadFiniteT}, finite nucleon mass \cite{FiniteMass}, neutrino heating by $e^\pm$ annihilations \cite{NeutrinoHeating}, QED plasma effects \cite{QEDPlasma} and more accurate numerical integration techniques \cite{BBNCode,PublicBBN1,PublicBBN2}.  In addition, the input reaction rates, including the neutron mean lifetime, have been more accurately and precisely measured and the uncertainties quantified (see e.g. \cite{BurlesNollett}).  The BBN code is one of the avatars of precision cosmology.  Moreover, the comparison of the predicted light-element abundances (D, $^4$He, $^3$He and $^7$Li) with their inferred primordial abundances is the earliest test of the hot big bang cosmology as well as a powerful cosmological probe of particle physics.

Recently, Sasankan et al \cite{Sasankan2018} have called attention to an effect which they claim changes the predictions for the abundances of the light elements significantly.  Specifically, they question the assumption that nuclei are in kinetic equilibrium with phase-space distributions that are well described by Maxwell-Boltzmann distributions.  They argue that kinetic equilibrium for nuclei in the relevant temperature range ($T\sim 1\,$MeV to $T\sim 0.05\,$MeV) is maintained by scattering with the semi-relativistic $e^\pm$ plasma; based on numerical simulations, they suggest that this leads to a distorted  kinetic distributions for nuclei, one that appears to be described by a MB distribution that is about 20\% hotter than the temperature of the relativistic plasma (see Fig.~1 in \cite{Sasankan2018}).  If their work is correct, this  leads to a large change in the predicted light-element abundances, which they compute.

The importance of BBN in cosmology motivates our work.  In this paper, we show that there is a deviation from kinetic equilibrium, but it is extremely small.  In particular, using the relativistic Boltzmann equation, we explicitly show that any distortion that arises from nuclei being thermalized by the EM plasma is of the order of the expansion rate over the scattering rate, or about $10^{-17}$, not 20\%. We discuss and quantify two related, small effects:  The lightest nuclei are slightly relativistic, $v \sim 10^{-4} - 10^{-3}$, and so there are corrections to the MB distribution of order 0.1\%. Further, because the nuclei are non-relativistic, in the absence of interactions with the electromagnetic (EM) plasma, their temperature would decrease faster than the electromagnetic plasma, and the continued transfer of a small amount of energy from the EM plasma to nuclei quickens the cooling of EM plasma.   This is a very tiny effect because the thermal energy of the EM plasma is a billion times greater than that of the kinetic energy carried by nuclei.

\section{Relativistic Boltzmann Equation}
The relativistic Boltzmann equation governing the phase space distribution of species $X$ in the RW expanding Universe is given by \cite{KolbTurner}:
\begin{eqnarray}
{\hat {\bf L}}[f_X] &=& {\hat {\bf C}}[f_X]
\end{eqnarray}
where the Liouville operator and collision term are given by
\begin{eqnarray}
 {\hat {\bf L}}[f_X] & = & E {\partial f_X \over \partial t} - H |{\bf p}|^2 {\partial f_X \over \partial E} \\
{\hat {\bf C}}[f_X] & = & -{1\over 2} \int d\Pi_a  d\Pi_i d\Pi_j  |{\cal M}|^2_{a + X \leftrightarrow i + j } (2\pi )^4\delta^{(4)}(\dots ) \nonumber \\
&\times& \left[f_a f_X (1\pm f_i)(1\pm f_j) -f_if_j (1\pm f_a)(1\pm f_X) \right] \label{C-term}
, \end{eqnarray}
 the $+$ is for bosons, the $-$ is for fermions, and $\hbar = k_B = c =1$ throughout.  Anticipating the problem of interest we have specialized to a single $2\leftrightarrow 2$ reaction.  More generally, the collision term should be summed over all possible interactions.
 
Next, we remind the reader that ${\hat {\bf C}}[f_X] = 0$ in the case that the particles (here $a$, $X$, $i$ and $j$) take on equilibrium distributions characterized by a temperature:
\beq \label{th-eq-dist}
f_{\rm EQ} = {1\over e^{(E-\mu )/T } \mp 1},
\eeq
with $\mu_a + \mu_X = \mu_i + \mu_j$.  That is, in the absence of expansion, the stationary solution, i.e., thermal equilibrium, is given by the usual FD (or BE) distributions for each species.  In the expanding Universe,  the growing scale factor $a(t)$ shifts the particle distributions through the effect of redshifting of particle momenta:  $|{\bf p}| \propto a^{-1}$, irrespective of mass, where $a(t)$ is the RW cosmic scale factor. Thus, in general, maintaining thermal distributions requires interactions that occur rapidly on the expansion timescale, $H^{-1}$.

\subsection{Collisionless, Nonrelativistic Limit}
In the nonrelativistic limit, the Liouville operator becomes
$${\hat {\bf L}} = M\left( {\partial \over \partial t} - 2HE_K {\partial \over \partial E_K} \right) ,$$
where $E_K = p^2/2M$. Further, for any phase space distribution $f(E_K, t)$ of the form $f= g(E_K = a^2{\tilde E})$, ${\hat {\bf L}}[f]$ vanishes.  This means that in the collisionless limit, the phase space distribution function $f$ simply evolves due to the redshifting of particle kinetic energy as $a^{-2}$.  If the initial phase-space distribution was thermal, then in the absence of collisions, the distribution remains thermal with a temperature that redshifts as $a^{-2}$.  This is a standard result.

More well known is that in the collisionless relativistic limit, 
$${\hat {\bf L }} = |{\bf p}| \left( {\partial \over \partial t} - 2H |{\bf p}| {\partial \over \partial E_K} \right) ,$$
an initially thermal distribution will remain thermal with a temperature that redshifts as $T \propto a^{-1}$, even in the absence of interactions.  The high precision to which the CMB is a blackbody spectrum today, 15\,Gyr after photon decoupling, gives strong testimony to the correctness of this result.

\subsection{Nucleon/nuclei heating by the EM plasma}
Around the time of BBN, the two constituents of the EM plasma, photons and $e^\pm$ pairs, have comparable abundances.  Since the $e^\pm$-scattering cross section with nucleons/nuclei is larger, that process is more important (see \cite{Sasankan2018} for a discussion of this point), and so we consider only the thermalizing reaction $e^\pm (p) + N (P) \leftrightarrow e^{\pm\prime }(p^\prime )+ N^\prime (P^\prime )$:
\begin{eqnarray}
 & &{\partial f_N (P) \over \partial t}  -  2HE_K {\partial f_N (P) \over \partial E_K}  =   -{1\over 2M} \int d\Pi_p  d\Pi_{p^\prime}d\Pi_{P^\prime}\nonumber \\
& & \times  |{\cal M}|^2_{e + N \leftrightarrow e^\prime+ N^\prime}  (2\pi )^4\delta^{(4)}(\dots ) [f_e (p) f_N (P) (1 - f_{e^\prime}) \nonumber \\
& & -   f_e(p^\prime )f_N (P^\prime ) (1 - f_e)] 
\label{B-eqn}
\end{eqnarray}
where $M$ is the mass of nuclide of interest, $d\Pi = d^3p/(2\pi )^3 2E$ is the usual LIPS, the matrix-element squared is
\bea
|{\cal M}|^2_{e + N \leftrightarrow e^\prime+ N^\prime}  = {16 Z^2 e^4 \over q^4} \left[ 4(p\cdot P)^2 + q^2 \left(m_e^2+M^2 + 2 p\cdot P \right) + \frac{q^4}2 \right] , \label{mel-sq}
\eea
$p^\mu$, $P^\mu$, $p'^\mu$, $P'^\mu$ are the four momenta of the particles, $e$, $N$, $e^\prime$, and $N^\prime$, and the momentum transfer is $q^\mu \equiv p^\mu-p'^\mu$.  Because  the quantum occupancy of nucleons/nuclei is small, we have neglected the Pauli blocking factors (for more about this, see Sec.~2.4). Here and in the following, we will often use $P,q$ to denote $|\vec P|, |\vec q|$.

The dimensions of the two sides of Eq.~\ref{B-eqn} are $[E] = [{\rm time}]^{-1}$, that of a rate.  By pulling out a factor of $\Gamma \equiv 4MZ^2e^4$ on the r.h.s., the remaining integral becomes dimensionless.  $\Gamma$ characterizes the interaction rate, and if we  compare it to the expansion rate of the Universe $H \simeq T^2/m_{\rm Pl}$,
$$ {\Gamma \over H} \simeq {Z^2e^4 m_{\rm Pl}M \over T^2} \sim {10^{21} \over (T/{\rm MeV})^2 } , $$
we see that the scattering rate of nuclei with thermal $e^\pm$ pairs is expected to be very high, of order $10^{21}$ scatterings per Hubble time, and so we expect any departures in the phase space distribution of nuclei from equilibrium to be very small, with size set by $H/\Gamma$.  We now explicitly show that this is indeed the case, though factors of $M/T$ raise the size of the departure slightly. In the ensuing section, we calculate the correct $M/T$ dependence.

\subsection{Perturbative estimate of non-equilibrium}
Because nuclei are non-relativistic, expansion cools their kinetic distribution faster than the EM plasma, driving a departure from kinetic equilibrium with the EM plasma ($1/a^2$ versus $1/a$).  However, elastic scatterings with the EM plasma heat the nuclei, and drive their distribution toward kinetic equilibrium with the plasma, at a rate $\Gamma \gg H$.   

To calculate the size of the expected small deviation from equilibrium, we write the distribution function for a nuclide as the equilibrium distribution plus a small correction: 
$ f_N (P)= f_{\rm EQ}(P) + \delta f (P)$. Applying the Liouville and collision operators to our {\it ansatz} and keeping the lowest-order terms in $\delta f$, we find:
\bea
{\hat{\bf L}}[f_N] & = & MH(E_K/T)f_{\rm EQ}   \nonumber  \\
{\hat{\bf C}}[f_N] & = & -\frac12 \int d \Pi_e d \Pi_{e'} d \Pi_{P'} |\cM|^2 (2\pi)^4 \delta^{(4)} (P^\mu + p^\mu - P'^\mu -p'^\mu) \times \nonumber\\ && \qquad \times \cbL f_e(p) [1-f_e(p')] \delta f(P) - f_e(p')[1-f_e(p)]\delta f(P') \cbR \nonumber \\
 & = & -\frac12 \int d \Pi_e d \Pi_{e'} d \Pi_{P'} |\cM|^2 (2\pi)^4 \delta^{(4)} (P^\mu + p^\mu - P'^\mu -p'^\mu) \times \nonumber\\ && \qquad \times f_e(p) [1-f_e(p')] \bL\delta f(P) - e^{(E_e -E_e')/T } \delta f(P') \bR \nonumber \\
 & \simeq & -\frac12 \int d \Pi_e d \Pi_{e'} d \Pi_{P'} |\cM|^2 (2\pi)^4 \delta^{(4)} (\dots) f_e(p) [1-f_e(p')] \times \nonumber\\ && \qquad \times \cbL \delta f(P) (E_e - E'_e)/T - \pL P- P' \pR \delta f'(P) \cbR \nonumber
 \eea
where in the final expression we have Taylor-expanded $e^{(E_e -E_e')/T}$ and $\delta f(P')$ and kept the lowest-order terms. The symbol $\delta f'(P)$ represents the partial derivative of $\delta f$ with respect to $P$.

The Boltzmann equation, ${\hat{\bf L}}[f_N] =  {\hat{\bf C}}[f_N]$, now leads to  an ordinary differential equation for $\delta f(P)$ whose coefficients are phase-space integrals that can computed numerically. However, we can obtain a parametric estimate for $\delta f$ by using the fact that the momentum transferred $q \sim T$ and the energy transferred $q^2/M$ are small compared to the nuclide mass $M$, so that 
\bea
\delta f(P)(E_e - E'_e)/T &\sim& {\cal O}(T/M) \delta f \nonumber
\\
\pL P- P' \pR \delta f'(P)   &\sim& {\cal O}[(T/M)^{1/2}] \delta f. \nonumber
\eea
The first term therefore enters at higher order in $T/M$. Using the matrix element from \Eq{mel-sq} and integrating over phase space, we find that the collision term is approximately
\beq
{\hat{\bf C}}[f_N] \sim \frac{32 \alpha^2 M^2 T^3 \ln(\theta_D/2) \cI(T)}{\pi |\vec P|^3} \delta f,
\eeq
where we have defined a dimensionless integral over the electron phase space,
$$\cI(T)=\int_{m_e/T}^\infty d \ep \, \ep^2 \sqrt{1-\frac{m_e^2}{T^2 \ep^2}} \frac{\exp\ep}{(\exp\ep+1)^2} \sim \cO(1),$$
and the Debye screening angle is $\theta_D\sim \alpha^{3/2}$. Dropping order one numbers, our parametric estimate for $\delta f$ is:
 \bea
{\delta f \over f_{\rm EQ}} \sim \frac H\Gamma \left( \frac MT\right)^{3/2} \frac1{\ln (\theta_D/2)} \sim {H \sqrt M \over \alpha^2  \ln(\theta_D/2) T^{3/2}}
  \sim 10^{-17} (T/\mev)^{1/2} 
.&& \eea
Thus, we have explicitly shown that the departure from kinetic equilibrium is tiny and characterized by $H/\Gamma$. Furthermore, since the collision term is dominated by the first derivative with respect to $P$, we can approximately determine the $P$-dependence of $\delta f(P)$ by integrating once over $P$. Our analysis above gives $\delta f'(P) \propto P^4 f_{\rm EQ}$, leading to
$$\delta f(P) \propto \pL 1 +  \frac{P^2}{3MT} \pR \frac{P}{\sqrt{2MT}} e^{- P^2/2MT} + \frac{\sqrt\pi}2 {\rm erfc}\pL \frac{P}{\sqrt{2MT}} \pR.$$
This gives the correct $P$-dependence to order $\sim \sqrt{T/M}$.

Sasankan et al.~\cite{Sasankan2018} describe in version 2 of their paper how they arrive at their result.  Starting with the nonrelativistic version of the Langevin equation and using ``the principle of equipartition of KE,'' they numerically simulate the thermalization of nuclei, specifically protons, at an EM plasma temperature of $T=0.1\mev$.  They find that protons have a kinetic distribution well described by a MB distribution at temperature of $T\simeq 0.12\mev$ (see Fig.~3), about 20\% warmer than the EM plasma.  This is consistent with the fact that at $T=0.1\mev$ the average KE of an electron or positron is 20\% greater than $1.5T$.

Their simulation gives results that are consistent with their incorrect assumption about equipartition of KE. In the nonrelativistic limit, thermal equilibrium implies equal KE for all particle species independent of mass; in the relativistic limit (or mixed nonrelativistic/relativistic limit) this is not true, cf., the average energy per particle for a boson is $2.7T$ and for a fermion is $3.15T$ compared to $1.5T$ for a nonrelativistic species.  While the Langevin and Fokker-Planck equations, which are used to describe the thermalization of heavy particles by lighter particles (e.g., Brownian motion), can be derived from the Boltzmann equations, the relativistic version of these equation is appropriate here.  Had they done this, they would not have needed their ``equipartition assumption'' and we believe they would have arrived at results consistent with ours.

\subsection{Relativistic correction to MB distribution}
Nuclei are slightly relativistic at the time of BBN:  $v^2 \sim T/M \sim 10^{-4} - 10^{-3}$ and thus the use of the MB distribution, $f (v) \propto v^2 \exp (-E_K/T)$, to describe their phase space distribution is not exact.  The correction is easy to compute by starting with the exact FD (or BE) distribution:
$${g_N \over e^{(E-\mu )/T}+1} \longrightarrow e^{-(E-\mu )/T} \propto e^{-E_K/T}, $$
where the first step follows from the fact $(E-\mu )/T \sim \ln (\eta^{-1}) + 3\ln (M/T) /2 \sim 25 \gg 1$ (which implies small phase-space occupancy) and $E_K \equiv E-M = (\gamma -1)M$.  Next, it is straightforward to show that 
$$p^2dp = M^3\gamma^5 v^2 dv .$$
Therefore, in the $E-\mu \gg T$ limit (relevant to cosmology), the fully relativistic phase space distribution is
\begin{equation}
{g_N \ p^2dp \over e^{(E-\mu )/T} +1 } \longrightarrow  g_N M^3\gamma^5 v^2 dv e^{-E_K/T}.
\end{equation}
Expanding $\gamma = 1/{(1-v^2)}^{1/2}$ and $e^{-E_K/T}$ in powers of $v^2$, we find the lowest-order correction to the MB distribution:
\begin{equation}
f_N (v) \propto  \left[ 1 + \left( {5\over 2} - {3\over 8}{Mv^2 \over T} \right) v^2 + {\cal O}(v^4) \right] v^2 dv e^{-Mv^2/2T}.
\end{equation}
The sign of the correction varies from positive, for $Mv^2 < 20T/3$, to negative, for $Mv^2 > 20T/3$.  While the thermal average of $Mv^2$ is $3T$, where the overall ${\cal O}(v^2)$ correction is still positive,  the Gamow peak for most of the important BBN nuclear reactions is at an energy of a few times the thermal average \cite{BurlesNollett}, where the correction can be negative.  In any case, the correction is small, of the order of 0.1\%, smaller than the experimental uncertainties in the nuclear reaction rates \cite{BurlesNollett}.  

\subsection{Nuclear kinetic drag}
Finally, we consider a distinct effect that has previously been ignored.   As discussed above, once nucleons and eventually nuclei become non-relativistic, for $T\ll1\,$GeV, their kinetic energies redshift as $a^{-2}$ and without heating their kinetic temperature would decrease as $a^{-2}$ as well.  The interactions of nuclei with the relativistic plasma heats the nuclei and keeps them in good thermal contact, as discussed above.  However, this heating depletes energy from the relativistic plasma and causes it to cool moderately faster than $1/a(t)$.  Using $dE = -pdV$ with
\begin{eqnarray}
E & = & a^3\, \rho \nonumber\\
\rho & = & \rho_{\rm EM} + {3\over 2}nT \nonumber \\
p &=& {1\over 3}\rho_{\rm EM} + nT \nonumber \\
V&=& a^3 \nonumber \\
\varepsilon & \equiv & {3nT/2 \over \rho_{\rm EM}} 
\end{eqnarray}
it is simple to show that 
$$ T \propto a^{-(1+\varepsilon /4)}$$
Here $n$ is the number density of nucleons/nuclei and $\varepsilon $, the ratio of the kinetic energy in nucleons/nuclei compared to the EM plasma, is approximately constant and equal to the one-eighth of the baryon-to-photon ratio $\eta$, or around $10^{-10}$.  Clearly this is a very tiny effect.  

By comparison, the annihilation of $e^\pm$ pairs ($T\sim 0.3\,$MeV to $T\sim 0.03\,$MeV) heats the photons.  The average slope of the temperature/scale factor relationship during this period is $T \propto a^{-0.84}$ rather than $a^{-1}$.  This much larger effect is incorporated into the standard BBN treatments.

\section{Conclusions}
It is often said that we are in the era of precision cosmology.  BBN, CMB last-scattering, and the evolution of CMB anisotropy are exemplars of such.  Both involve precision calculations based on well-understood physics, and both have a history that traces back to the discovery of the CMB more than 50 years ago.  CMB anisotropies have been computed with a theoretical uncertainty of less than 0.1\% and have been measured to the cosmic variance limit for multipoles up to $\sim 2000$.  The estimated theoretical uncertainty in the BBN code for computing $^4$He is less than 0.1\% with the uncertainty in the neutron lifetime adding a similar amount to the error budget \cite{BBNCode}.  The theoretical uncertainties for the other light-element abundances are at a similar level \cite{BurlesNollett}.  Moreover, the precision determination of the baryon density links the two: BBN and CMB each separately determine the baryon density to per cent level, and the two determinations agree \cite{BaryonDensity}.

This backdrop of precision cosmology made the claim of a 20\% correction to the kinetic distribution functions of nuclei \cite{Sasankan2018} of potential great importance and motivated our work.  To wit, we have solved the relativistic Boltzmann equation for the nuclear phase-space distribution and explicitly shown that any non-equilibrium effect arising due to $e^\pm$ (and photon) scattering with nuclei is many orders-of-magnitude smaller than this, owing to the very large scattering rate compared to the expansion rate. We have not been able to identify the source of the discrepancy with \cite{Sasankan2018}.

Finally, we identified two new small effects:  relativistic corrections to the MB distributions for nuclei and a nuclear kinetic drag on the EM plasma which hastens its cooling.  The latter of these effects is extremely tiny.  The former, the relativistic corrections, are expected to be of the order of 0.1\% and their effect on the light-element abundances is expected to be similar, but has yet to be computed.

\vskip 20pt

We thank Nikita Blinov for very helpful discussions and Susan Gardner and Grant Mathews for correspondence. This work was supported in part by the Kavli Institute for Cosmological Physics at the University of Chicago through grant NSF PHY-1125897 and an endowment from the Kavli Foundation and its founder Fred Kavli. SDM is an employee of Fermilab, operated by Fermi Research Alliance, LLC under Contract No.~De-AC02-07CH11359 with the United States Department of Energy.

\end{document}